\documentclass[preprint,aps,12pt,showpacs,nofootinbib,tightenlines]{revtex4}
\usepackage{amsmath}
\usepackage{amssymb}
\usepackage{epsfig}
\usepackage{graphicx}
\begin{document}


\newcommand{\beq}{\begin{eqnarray}}
\newcommand{\eeq}{\end{eqnarray}}
\newcommand{\non}{\nonumber\\ }

\newcommand{\kst}{ K_0^*(1430) }
\newcommand{\kstb}{ \bar{K}_0^*(1430)^0 }
\newcommand{\kt}{ K_0^* }

\newcommand{\etap}{\eta^{(\prime)} }
\newcommand{\etapr}{\eta^\prime }
\newcommand{\pe}{\phi_{\eta}^A}
\newcommand{\pepr}{\phi_{\eta'}^A}
\newcommand{\pep}{\phi_{\eta}^P}
\newcommand{\peprp}{\phi_{\eta'}^P}
\newcommand{\pet}{\phi_{\eta}^T}
\newcommand{\peprt}{\phi_{\eta'}^T}
\newcommand{\fpi}{f_{\pi} }
\newcommand{\feta}{f_{\eta} }
\newcommand{\fetap}{f_{\eta'} }
\newcommand{\re}{r_{\eta} }
\newcommand{\rep}{r_{S} }
\newcommand{\mb}{m_{B_s} }
\newcommand{\moe}{m_{0\eta} }
\newcommand{\moep}{m_{0\eta'} }

\newcommand{\psl}{ p \hspace{-1.8truemm}/ }
\newcommand{\nsl}{ n \hspace{-2.2truemm}/ }
\newcommand{\vsl}{ v \hspace{-2.2truemm}/ }
\newcommand{\epsl}{\epsilon \hspace{-1.8truemm}/\,  }

\def \epjc{ Eur. Phys. J. C}
\def \jpg{  J. Phys. G}
\def \npb{  Nucl. Phys. B}
\def \plb{  Phys. Lett. B}
\def \pr{  Phys. Rep.}
\def \prd{  Phys. Rev. D}
\def \prl{  Phys. Rev. Lett.}
\def \zpc{  Z. Phys. C}
\def \jhep{ J. High Energy Phys.}
\def \rmp{ Rev. Mod. Phys.}

\title{$B \to \kst \eta^{(\prime)}$ decays in the pQCD approach
\footnote{This work is partially supported by the National Natural Science
Foundation of China under Grant No.10575052, 10605012 and¡¡10735080.}}
\author{Xin Liu$^{a}$  \footnote{js.xin.liu@gmail.com}, Zhi-Qing Zhang$^{a,b}$ and
 Zhen-Jun Xiao$^a$ \footnote{ xiaozhenjun@njnu.edu.cn}}
\affiliation{{\it $a$ Department of Physics and Institute of
Theoretical Physics, Nanjing Normal University, Nanjing, Jiangsu
210097, P.R.China}\\ {\it $b$ Department of Mathematics and Physics,
Henan University of Technology, Zhengzhou, Henan 450052, P.R.China
}} 
\date{\today}
\begin{abstract}
Based on the assumption of two-quark structure of the scalar meson
$\kst$, we calculate the CP-averaged branching ratios for $B \to
\kst \etap$ decays in the framework of the perturbative QCD (pQCD)
approach here. We perform the evaluations in two scenarios for the
scalar meson spectrum. We find that: (a) the pQCD predictions for
$Br(B \to \kst \eta^{(\prime)})$ which are about $10^{-5}-10^{-6}$,
basically agree with the data within large theoretical uncertainty;
(b) the agreement between the pQCD predictions and the data in
Scenario I is better than that in Scenario II, which can be tested
by the forthcoming LHC experiments; (c) the annihilation
contributions play an important role for these considered decays.
\end{abstract}


\maketitle


Very recently, the branching ratios of $B \to {\kst} \eta$
decays have been measured by BaBar collaboration~\cite{prl97}
with good precision:
\beq
Br(B^+\to {\kt}^+(1430) \eta) &=& 18.2 \pm 2.6 \pm 2.6 \times 10^{-6}\;,\non
 Br(B^0\to {\kt}^0(1430) \eta) &=& 11.0 \pm 1.6 \pm 1.5 \times
 10^{-6}\;.
 \label{eq:exp1}
\eeq

It is well-known that the underlying structure of scalar mesons is not
well established theoretically (for a review, see e.g.~\cite{plb667,rmp71,jpg28}).
Presently, motivated by the large
number of $B$ production and decay events expected at the
forthcoming LHC experiments, the scalar meson spectrum is becoming
one of the interesting topics for both experimental and theoretical
studies. It is hoped that through the study of $B \to S
P$(S and P are scalar and pseudoscalar mesons) decays,
old puzzles related to the internal structure and related parameters, e.g., the masses
and widths, of light scalar mesons can receive new understanding. On
one hand, $B \to SP$ is another window to study their
properties~\cite{CCY06,DLR06}; on the other hand, CP asymmetries of
these decays provide another way to measure the CKM angles $\beta$
and maybe $\alpha$~\cite{SLSD}. Additionally, $B \to SP$ decays have
to be taken into account in order to analyze the $B \to 3P$ decays
in the different channels~\cite{BFKLL} and perhaps these decays can
be used to study new physics(NP) effects~\cite{GMM}.

At present, some $B\to S P, S V$ decays~\cite{CCY06,SWL07,CCY08}
have been studied, for example, by employing the QCD factorization
(QCDF) approach~\cite{bbns99} or the perturbative QCD (pQCD)
approach~\cite{cl97,li2003,lb80}. In this paper, based on the
assumption of two-quark structure of scalar $\kt$ meson (For the
sake of simplicity, we will use $\kt$ to denote $\kst$ in the
following section), we will calculate the branching ratios for the
four $B \to {\kt}^+ \eta, {\kt}^+ \etapr, {\kt}^0 \eta$ and ${\kt}^0
\etapr$ decays by employing the  pQCD factorization approach.

This paper is organized as follows. In
Sec.~\ref{sec:f-work}, we calculate analytically the related Feynman
diagrams and present the various decay amplitudes for the studied
decay modes. In Sec.~\ref{sec:n-d}, we show the numerical results
for the branching ratios of $B \to \kt \eta^{(\prime)}$ decays. A
short summary and some phenomenological discussions are also
included in this section.

\section{Perturbative calculations}\label{sec:f-work}

Since the b quark is rather heavy we consider the $B$ meson at rest
for simplicity. It is convenient to use light-cone coordinate $(p^+,
p^-, {\bf p}_T)$ to describe the meson's momenta. Using the
light-cone coordinates the $B$ meson and the two final state meson
momenta can be written as \beq P_1 =\frac{M_{B}}{\sqrt{2}} (1,1,{\bf
0}_T),\quad P_2 =\frac{M_{B}}{\sqrt{2}} (1,0,{\bf 0}_T), \quad P_3
=\frac{M_{B}}{\sqrt{2}} (0,1,{\bf 0}_T), \eeq respectively, here the
light meson masses have been neglected. Putting the light (anti-)
quark momenta in $B$, $\eta$ and $\kt$ mesons as $k_1$, $k_2$, and
$k_3$, respectively, we can choose \beq k_1 = (x_1 P_1^+,0,{\bf
k}_{1T}), \quad k_2 = (x_2 P_2^+,0,{\bf k}_{2T}), \quad k_3 = (0,x_3
P_3^-,{\bf k}_{3T}). \eeq Then, after the integration over $k_1^-$,
$k_2^-$, and $k_3^+$,  the decay amplitude for $B^+ \to {\kt}^+
\eta$ decay, for example, can be conceptually written as \beq {\cal
A}(B^+ \to {\kt}^+ \eta) &\sim &\int\!\! d x_1 d x_2 d x_3 b_1 d b_1
b_2 d b_2 b_3 d b_3 \non && \cdot \mathrm{Tr} \left [ C(t)
\Phi_{B}(x_1,b_1) \Phi_{\eta}(x_2,b_2) \Phi_{\kt}(x_3, b_3) H(x_i,
b_i, t) S_t(x_i)\, e^{-S(t)} \right ], \label{eq:a2} \eeq where
$k_i$ are the momenta of light quarks included in each meson, the
term $\mathrm{Tr}$ denotes the trace over Dirac and color indices,
$C(t)$ is the Wilson coefficient evaluated at scale $t$, the hard
kernel $H(k_1,k_2,k_3,t)$ is the hard part and can be calculated
perturbatively, the function $\Phi_M$ is the wave function, the
function $S_t(x_i)$ describes the threshold resummation ~\cite{li02}
which smears the end-point singularities on $x_i$, and the last
term, $e^{-S(t)}$, is the Sudakov form factor which suppresses the
soft dynamics effectively.

For the two-body charmless $B$ meson decays, the related weak
effective Hamiltonian $H_{eff}$ can be written as \cite{buras96}
\beq \label{eq:heff} {\cal H}_{eff} = \frac{G_{F}} {\sqrt{2}} \,
\left[ V_{ub}^* V_{us} \left (C_1(\mu) O_1^u(\mu) + C_2(\mu)
O_2^u(\mu) \right) - V_{tb}^* V_{ts} \, \sum_{i=3}^{10} C_{i}(\mu)
\,O_i(\mu) \right] \; , \eeq where $C_i(\mu)$ are the Wilson
coefficients at the renormalization scale $\mu$ and $O_i$ are the
four-fermion operators for the case of $\bar b \to \bar s $
transition \cite{buras96}. For the Wilson coefficients $C_i(\mu)$
($i=1,\ldots,10$), we will use the leading order (LO) expressions,
although the next-to-leading order (NLO)  results already exist in
the literature ~\cite{buras96}. This is the consistent way to cancel
the explicit $\mu$ dependence in the theoretical formulae. For the
renormalization group evolution of the Wilson coefficients from
higher scale to lower scale, we use the formulae as given in
Ref.\cite{luy01} directly.

In the two-quark picture, the decay constants $f_{\kt}$ and $\bar
f_{\kt}$ for a scalar meson $\kt$ are defined by:
\begin{eqnarray}
\langle {\kt}(p)|\bar{q}_2 \gamma_{\mu}q_1 |0
\rangle=f_{\kt}p_{\mu},\,\,\, \langle {\kt}(p)|\bar{q}_2 q_1 |0
\rangle=m_{\kt}\bar{f}_{\kt} ,
\end{eqnarray}
where $m_{\kt}(p)$ is the mass (momentum) of the scalar meson, and
\beq f_{\kt}&=&-0.025 \pm 0.002 {\rm GeV},\quad
\bar{f}_{\kt}=-0.300\pm 0.030 {\rm Gev}\eeq in Scenario I,  and \beq
f_{\kt}&=&0.037 \pm 0.004 {\rm GeV}, \quad \bar{f}_{\kt}=0.445\pm
0.050{\rm  Gev}, \eeq in Scenario II~\cite{CCY06}, respectively.

The light-cone wave function 
of the scalar meson $\kt$ is defined as: \beq
\Phi_{\kt,\alpha\beta}&=&\frac{i}{\sqrt{2 N_C}}\bigg\{ \psl
\phi_{\kt}(x)+ m_{\kt}\phi^S_{\kt}(x)+
m_{\kt}(\vsl\nsl-1)\phi^T_{\kt}(x)\bigg\}_{\alpha\beta} \eeq where
$v=(0,1,{\bf 0}_T)$ and $n=(1,0,{\bf 0}_T)$ are the dimensionless
light-like unit vectors.

The twist-2 light-cone distribution amplitude $\phi_{\kt}(x,\mu)$
can be expanded as the Gegenbauer polynomials: \beq
\phi_{\kt}(x,\mu)&=&\frac{3}{\sqrt{2N_c}}x(1-x)\biggl\{f_{\kt}(\mu)+\bar
f_{\kt}(\mu)\sum_{m=1}^\infty B_m(\mu)C^{3/2}_m(2x-1)\biggr\}, \eeq
where the values for Gegenbauer moments are taken at scale $\mu=1
\mbox{GeV}$~\cite{CCY06}:
$B_1=0.58\pm0.07$,$B_3=-1.20\pm0.08$(Scenario I) and
 $B_1=-0.57\pm0.13$,$B_3=-0.42\pm0.22$(Scenario II).

As for the twist-3 distribution amplitudes $\phi_{\kt}^S$ and
$\phi_{\kt}^T$, we adopt the asymptotic form:
\beq
\phi^S_{\kt}&=& \frac{1}{2\sqrt {2N_c}}\bar f_{\kt},\,\,\,\,\,\,\,\phi_{\kt}^T=
\frac{1}{2\sqrt {2N_c}}\bar f_{\kt}(1-2x).
\eeq


The B meson is treated as a heavy-light system. We here use the same
B meson wave function as in Ref.~\cite{liu06,guodq07}. For the
$\eta-\eta'$ system, we use the quark-flavor basis with $\eta_q =
(u\bar u +d\bar d)/\sqrt{2} $ and $\eta_s=s\bar s$, employ the same
wave function, the identical distribution amplitudes
$\phi_{\eta_{q,s}}^{A,P,T}$, and use the same values for other
relevant input parameters, such as $f_q=(1.07\pm0.02)f_{\pi}$,
$f_s=(1.34\pm0.06)f_{\pi}$, $\phi=39.3^\circ \pm 1.0^\circ $, etc.,
as given in Ref.~\cite{feldmann}. From those currently known
studies\cite{liu06,guodq07,ligluon}  we believe that there is no
large room left for the contribution due to the gluonic component of
$\etap$, and therefore neglect the possible gluonic component in
both $\eta$ and $\eta'$ meson.

We firstly  take  $B^+ \to {\kt}^+ \eta$ decay mode as an example,
and then extend our study to $B^+ \to {\kt}^+ \eta'$ and $ B^0 \to
{\kt}^0 \etap$ decays. Similar to the leading order $B \to K \etap$
decays in Ref.~\cite{xiao08a}, there are 8 types of diagrams
contributing to the $B^+ \to {\kt}^+ \eta $ decays, as illustrated
in Fig.1. We first calculate the usual factorizable diagrams (a) and
(b). Operators $O_{1-4,9,10}$ are $(V-A)(V-A)$ currents, the sum of
their amplitudes is given as \beq F_{e\kt}&=& 8 \pi C_F
m_{B}^2\int_0^1 d x_{1} dx_{3}\, \int_{0}^{\infty} b_1 db_1 b_3
db_3\, \phi_{B}(x_1,b_1)\left\{ h_e(x_1,x_3,b_1,b_3)\;E_e(t_a)
\right.\non & &\left. \cdot \left[(1+x_3 )\phi_{\kt}(x_3) + r_S (1-2
x_3) (\phi^S_{\kt}(x_3)+\phi^T_{\kt}(x_3))\right]  \right.\non & &
\left. + 2 r_S \phi^S_{\kt} (x_3)\; h_e(x_3,x_1,b_3,b_1) \;E_e(t_b)
\right\}. \label{eq:ab} \eeq where $\rep=m_{\kt}/m_B$; $C_F=4/3$ is
a color factor.

\begin{figure}[t,b]
\vspace{0.3 cm} \centerline{\epsfxsize=12 cm \epsffile{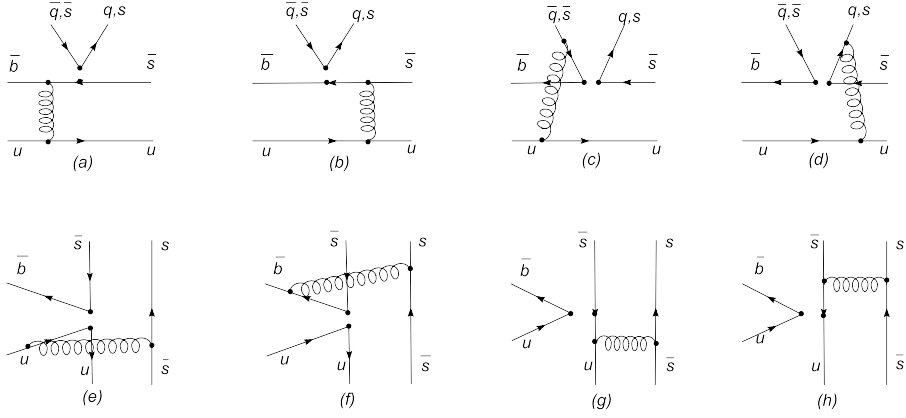}}
\vspace{0.4cm} \caption{ Typical Feynman diagrams contributing to
the $B^+ \to {\kt}^+ \eta$  decays, where diagrams (a) and (b)
contribute to the $B \to \kt$ form factor $F_{0,1}^{B \to \kt}$.}
\label{fig:fig1}
\end{figure}

The contributions from the operators $O_{5,6,7,8}$ can be written as
\beq F_{e\kt}^{P1}&=&-F_{e\kt} \; .\\ F_{e\kt}^{P2}&=& 16\pi C_F
m_{B}^2 \re \int_0^1 d x_{1} dx_{3}\, \int_{0}^{\infty} b_1 db_1 b_3
db_3\, \phi_{B}(x_1,b_1) \non && \cdot \left\{
\left[\phi_{\kt}(x_3)+ r_S [(2+x_3) \phi^S_{\kt}(x_3)- x_3
\phi^T_{\kt}(x_3)]\right] \; h_e(x_1,x_3,b_1,b_3)\; E_e(t_a)\right.
\non && \left. + 2 r_S \phi^S_{\kt} (x_3) \; h_e(x_3,x_1,b_3,b_1)\;
E_e(t_b)\right\}\; , \eeq where $\re=m_0^{\eta_{q}}/m_{B}$ and/or
$\re=m_0^{\eta_{s}}/m_{B}$.

For the hard spectator diagrams 1(c) and 1(d), the corresponding
decay amplitudes can be written as
\beq
M_{e\kt}&=&
\frac{32}{\sqrt{6}}\pi C_F m_{B}^2 \int_{0}^{1}d x_{1}d x_{2}\,d
x_{3}\,\int_{0}^{\infty} b_1d b_1 b_2d b_2\, \phi_{B}(x_1,b_1)
\phi_{\eta}^A(x_2)\left\{ \left[(1-x_2)\phi_{\kt}(x_3)\right.\right.
\non & &\left.\left.
 - r_S x_3 (\phi^S_{\kt}(x_3)-
\phi^T_{\kt}(x_3))\right]E_{ne}(t_c) h_{ne}^c(x_1,x_2,x_3,b_1,b_2)
-h_{ne}^d(x_1,x_2,x_3,b_1,b_2)\right.\non & & \left. \cdot
\left[(x_2+x_3)\phi_{\kt}(x_3) - r_S x_3 (\phi^S_{\kt}(x_3)+
\phi^T_{\kt}(x_3))\right] E_{ne}(t_d)
\right\} \; , \\
M_{e\kt}^{P1}&=&-\frac{32}{\sqrt{6}}\pi C_F m_{B}^2 \int_{0}^{1}d
x_{1}d x_{2}\,d x_{3}\,\int_{0}^{\infty} b_1d b_1 b_2d b_2\,
\phi_{B}(x_1,b_1) \re \left\{
\left[(1-x_2)\phi_{\kt}(x_3)\right.\right.\non
 & & \left.\left.\cdot (\phi_\eta^P(x_2)
+\phi_\eta^T(x_2)) - r_S \left(\phi_\eta^P(x_2)[(1-x_2+x_3)
\phi^S_{\kt}(x_3)-(1-x_2-x_3)\phi^T_{\kt}(x_3)]\right.
\right.\right.\non && \left.\left.\left. +\phi_\eta^T(x_2)
[(1-x_2-x_3)\phi_{\kt}^S(x_3)-(1-x_2+x_3)\phi_{\kt}^T]\right)\right]
h_{ne}^c(x_1,x_2,x_3,b_1,b_2)E_{ne}(t_c)\right.\non &&\left. -
\left[x_2(\phi_\eta^P(x_2)-\phi_\eta^T(x_2))\phi_{\kt}(x_3)+ r_S
(x_2 (\phi_\eta^P(x_2)-\phi_\eta^T(x_2))(\phi^S_{\kt}(x_3)-
\phi^T_{\kt}(x_3))\right.\right. \non &&\left.\left. + x_3
(\phi_\eta^P(x_2)+\phi_\eta^T(x_2))(\phi^S_{\kt}(x_3)+
\phi^T_{\kt}(x_3)))\right] E_{ne}(t_d)h_{ne}^d(x_1,x_2,x_3,b_1,b_2)
\right\} \;, \\ M_{e\kt}^{P2}&=& -\frac{32}{\sqrt{6}}\pi C_F m_{B}^2
\int_{0}^{1}d x_{1}d x_{2}\,d x_{3}\,\int_{0}^{\infty} b_1d b_1 b_2d
b_2\, \phi_{B}(x_1,b_1) \phi_{\eta}^A(x_2)\left\{ E_{ne}(t_c)
\right. \non & &  \left. \cdot \left[(1-x_2+x_3)\phi_{\kt}(x_3) -
r_S x_3 (\phi^S_{\kt}(x_3)+
\phi^T_{\kt}(x_3))\right]h_{ne}^c(x_1,x_2,x_3,b_1,b_2) \right. \non
& &\left.   - \left[ x_2 \phi_{\kt}(x_3) - r_S x_3
(\phi^S_{\kt}(x_3)- \phi^T_{\kt}(x_3))\right]
E_{ne}(t_d)h_{ne}^d(x_1,x_2,x_3,b_1,b_2) \right\} \; . \eeq

For the non-factorizable annihilation diagrams 1(e) and 1(f), we
find
\beq
M_{a\kt}&=& \frac{32}{\sqrt{6}}\pi C_F m_{B}^2
\int_{0}^{1}d x_{1}d x_{2}\,d x_{3}\,\int_{0}^{\infty} b_1d b_1 b_2d
b_2\, \phi_{B}(x_1,b_1)\left\{
\left[(1-x_3)\phi_\eta^A(x_2)\phi_{\kt}(x_3) \right.\right. \non & &
\left.\left.
 - \re r_S \left(\phi_\eta^P(x_2)[(1+x_2-x_3)
\phi^S_{\kt}(x_3)-
(1-x_2-x_3)\phi^T_{\kt}(x_3)]+\phi_\eta^T(x_2)\right.\right.\right.
\non && \left.\left.\left.
[(1-x_2-x_3)\phi_{\kt}^S(x_3)-(1+x_2-x_3)\phi_{\kt}^T(x_3)]
\right)\right] E_{na}(t_e) h_{na}^e(x_1,x_2,x_3,b_1,b_2) \right.\non
& & \left.- \left[x_2 \phi_\eta^A(x_2)\phi_{\kt}(x_3)- \re   r_S
\left(\phi_\eta^P(x_2)[(3+x_2-x_3) \phi^S_{\kt}(x_3)+(1-x_2-x_3)
\right.\right. \right.\non && \left. \left.\left. \cdot
\phi^T_{\kt}(x_3)]+\phi_\eta^T(x_2)
[(-1+x_2+x_3)\phi_{\kt}^S(x_3)+(1-x_2+x_3)\phi_{\kt}^T(x_3)]\right)\right]
 \right. \non && \left.
 \cdot
E_{na}(t_f)h_{na}^f(x_1,x_2,x_3,b_1,b_2) \right\} \;,
 \eeq
 \beq
M_{a\kt}^{P1}&=& -\frac{32}{\sqrt{6}}\pi C_F m_{B}^2 \int_{0}^{1}d
x_{1}d x_{2}\,d x_{3}\,\int_{0}^{\infty} b_1d b_1 b_2d b_2\,
\phi_{B}(x_1,b_1)\left\{ \left[ \re
x_2\phi_{\kt}(x_3)(\phi_\eta^P(x_2) \right.\right. \non
 & & \left.\left.
+\phi_\eta^T(x_2)) + r_S (1 -x_3)\phi_\eta^A(x_2)(
\phi^S_{\kt}(x_3)- \phi^T_{\kt}(x_3))\right]
 E_{na}(t_e)h_{na}^e(x_1,x_2,x_3,b_1,b_2)\right. \non & & \left.
 + \left[ \re
(2-x_2)(\phi_\eta^P(x_2)+\phi_\eta^T(x_2))
 \phi_{\kt}(x_3)+r_S (1+x_3)(\phi^S_{\kt}(x_3)-
 \phi^T_{\kt}(x_3))\right.\right.\non && \left.\left.\cdot
\phi_\eta^A(x_2)\right]E_{na}(t_f)h_{na}^f(x_1,x_2,x_3,b_1,b_2)
\right\} \; ,
\eeq

For the factorizable annihilation diagrams 1(g) and 1(h), we have
\beq F_{a\kt}&=&  8 f_B \pi C_F m_{B}^2\int_0^1 d x_{2} dx_{3}\,
\int_{0}^{\infty} b_2 db_2 b_3 db_3\, \left\{ \left[ x_2
\phi_\eta^A(x_2) \phi_{\kt}(x_3)-2\re r_S\right.\right. \non & &
\left.\left.  \cdot  \left((x_2 +
1)\phi^P_{\eta}(x_2)+(x_2-1)\phi^T_{\eta}(x_2)\right)\phi_{\kt}^S(x_3)\right]
 h_a(x_2,1-x_3,b_2,b_3) \right. \non && \left.\cdot
E_a(t_g)  + \left[(x_3-1)\phi_\eta^A(x_2) \phi_{\kt}(x_3)-2 \re r_S
\phi_\eta^P(x_2) \left((x_3-2)\phi^S_{\kt} (x_3)
\right.\right.\right.\non && \left.\left.\left.- x_3
\phi_{\kt}^T(x_3)\right) \right] E_a(t_h)h_a(1-x_3,x_2,b_3,b_2)
\right\} \\ F_{a\kt}^{P2}&=& 16 f_B \pi C_F m_{B}^2  \int_0^1 d
x_{2} dx_{3}\, \int_{0}^{\infty} b_2 db_2 b_3 db_3\,\left\{ \left[2
r_S \phi_\eta^A(x_2) \phi^S_{\kt}(x_3) \right.\right. \non & &
\left.\left.-\re
 x_2 (\phi_\eta^P(x_2)- \phi_\eta^T(x_2))\phi_{\kt}(x_3) \right]
h_a(x_2,1-x_3,b_2,b_3) E_a(t_g)\right. \non && \left.  + \left[ r_S
(1-x_3) \phi_\eta^A(x_2) (\phi_{\kt}^S(x_3)+\phi_{\kt}^T(x_3))- 2
\re \phi_\eta^P(x_2)\phi_{\kt} (x_3) \right]
 \right. \non && \left. \cdot
E_a(t_h)h_a(1-x_3,x_2,b_3,b_2) \right\}\;.
 \eeq

For the $B^+ \to {\kt}^+ \eta$ decay, besides the Feynman diagrams
as shown in Fig.~1 where the upper emitted meson is the $\eta$, the
Feynman diagrams obtained by exchanging the position of ${\kt}^+$
and $\eta$ also contribute to this decay mode. The decays
amplitudes for the first four new diagrams can be obtained
by the replacements
\beq
\phi_{\kt}\longleftrightarrow \pe, \quad
\phi^S_{\kt}\longleftrightarrow\pep , \quad
\phi^T_{\kt}\longleftrightarrow\pet , \quad
r_S\longleftrightarrow\re .
\eeq

For the last four annihilation diagrams, the decay
amplitudes can be written as,
\beq
M_{a\eta}&=& \frac{32}{\sqrt{6}}\pi C_F m_{B}^2
\int_{0}^{1}d x_{1}d x_{2}\,d x_{3}\,\int_{0}^{\infty} b_1d b_1 b_2d
b_2\, \phi_{B}(x_1,b_1)\biggl\{
\biggl[(1-x_3)\phi_{\kt}(x_2)\pe(x_3)  \non
 & &
 + \re r_S \left(\phi_{\kt}^S(x_2)[(1+x_2-x_3) \pep(x_3)-
(1-x_2-x_3)\pet(x_3)]+\phi_{\kt}^T(x_2) [\pep(x_3)\right. \non &&
\left. \cdot (1-x_2-x_3)-(1+x_2-x_3)\pet(x_3)]
\right)\biggr]h_{na}^e(x_1,x_2,x_3,b_1,b_2)E_{na}(t_e) -
E_{na}(t_f)\non &&\cdot
 \biggl[x_2
\phi_{\kt}(x_2)\phi_\eta^A(x_3)+ \re r_S
\left(\phi_{\kt}^S(x_2)[(3+x_2-x_3) \pep(x_3)+
(1-x_2-x_3)\pet(x_3)]\right. \non && \left.
-\phi_{\kt}^T(x_2)[(1-x_2-x_3)\pep(x_3)-(1-x_2+x_3)\pet(x_3)]\right)\biggr]h_{na}^f(x_1,x_2,x_3,b_1,b_2)
 \biggr\} \;,
 \eeq   \beq
M_{a\eta}^{P1}&=& \frac{32}{\sqrt{6}}\pi C_F m_{B}^2 \int_{0}^{1}d
x_{1}d x_{2}\,d x_{3}\,\int_{0}^{\infty} b_1d b_1 b_2d b_2\,
\phi_{B}(x_1,b_1)\left\{ \left[ r_S
x_2(\phi_{\kt}^S(x_2)+\phi_{\kt}^T(x_2)) \right.\right. \non
 & & \left.\left.\cdot
\pe(x_3) - \re (1 -x_3)\phi_{\kt}(x_2)( \pep(x_3)- \pet(x_3))\right]
 E_{na}(t_e)h_{na}^e(x_1,x_2,x_3,b_1,b_2)+\right. \non & & \left.
  \left[r_S
(2-x_2)(\phi_{\kt}^S(x_2)+\phi_{\kt}^T(x_2)) \pe(x_3)-\re
(1+x_3)\phi_{\kt}(x_2)(\pep(x_3)-
 \pet(x_3))\right]\right.\non && \left.\cdot
E_{na}(t_f)h_{na}^f(x_1,x_2,x_3,b_1,b_2) \right\} \; , \\
F_{a\eta}&=& 8 f_B \pi C_F m_{B}^2\int_0^1 d x_{2} dx_{3}\,
\int_{0}^{\infty} b_2 db_2 b_3 db_3\, \left\{ \left[ x_2
\phi_{\kt}(x_2)\phi_\eta^A(x_3)+2 \re r_S \right.\right. \non & &
\left.\left. \cdot \left((x_2 +
1)\phi^S_{\kt}(x_2)+(x_2-1)\phi^T_{\kt}(x_2)\right)\pep(x_3)\right]
 E_a(t_g)h_a(x_2,1-x_3,b_2,b_3) \right. \non && \left.
-\left[(1-x_3)\phi_{\kt}(x_2)\phi_\eta^A(x_3) +2 \re r_S
\phi_{\kt}^S(x_2) \left((2-x_3)\pep(x_3) + x_3 \pet(x_3)\right)
\right]\right. \non && \left. \cdot E_a(t_h)h_a(1-x_3,x_2,b_3,b_2)
\right\} \\ F_{a\eta}^{P2}&=& 16 f_B \pi C_F m_{B}^2 \int_0^1 d
x_{2} dx_{3}\, \int_{0}^{\infty} b_2 db_2 b_3 db_3\,\left\{ \left[2
\re \phi_{\kt}(x_2) \pep(x_3)- r_S x_2\pe(x_3)\right.\right. \non &
& \left.\left.\cdot
 (\phi_{\kt}^S(x_2)- \phi_{\kt}^T(x_2)) \right]
h_a(x_2,1-x_3,b_2,b_3) E_a(t_g)+ h_a(1-x_3,x_2,b_3,b_2)\right. \non
&& \left. \cdot E_a(t_h) \left[2 r_S \phi_{\kt}^S(x_2)
\phi_\eta^A(x_3) + \re (1-x_3)\phi_{\kt} (x_2)(\pep(x_3)+\pet(x_3))
\right]
 \right\}\;.
 \eeq
The explicit expressions of hard functions $E_{e,ne;na,a}(t_i)$ and
$h_{e,ne;na,a}(x_i,b_j), \cdots$ can be found for example in
Ref.\cite{liu06,guodq07,xiao08a}.

Before writing the total amplitude of $B^+\to {\kt}^+ \eta$ decay,
we firstly define the combinations of Wilson coefficients as
usual~\cite{AKL98},
\beq a_1&=&C_2+C_1/3,\quad a_2=C_1+C_2/3,\non
a_i&=& C_i+C_{i\pm 1}/3,\quad  i=3-10.\eeq where the upper (lower)
sign applies, when $i$ is odd (even).

By combining the contributions from different diagrams, the total
decay amplitudes for $B^+ \to {\kt}^+ \eta$, for example, can be
written as {\small \beq {\cal M}({\kt}^+ \eta)&=& \zeta_q
F_{e\kt}f_q \left\{ \lambda_u a_2 - \lambda_t
\left[2(a_3-a_5)-\frac{1}{2}(a_7-a_9)\right]\right\} - \zeta_s f_s
\lambda_t \left\{F_{e\kt}\right.\non && \left.\cdot \left[a_3 +a_4 -
a_5+\frac{1}{2}(a_7-a_9-a_{10})\right]+ F_{e\kt}^{P2} \left(
a_6-\frac{1}{2}a_8\right) \right\} + \zeta_q\left\{
M_{e\kt}\right.\non &&\left. \cdot\left[ \lambda_u C_2 - \lambda_t
(2 C_4 + \frac{1}{2}C_{10})\right]-M_{e\kt}^{P2} \lambda_t (2 C_6 +
\frac{1}{2}C_8)\right\} - \zeta_s \lambda_t\left\{ M_{e\kt}\right.
\non &&\left.\cdot (C_3+C_4-\frac{1}{2}(C_9+C_{10})) + M_{e\kt}^{P1}
(C_5-\frac{1}{2}C_7)+ M_{e\kt}^{P2} (C_6-\frac{1}{2}C_8)\right\}
\non && +\zeta_s \left\{ M_{a\kt} \left[ \lambda_u C_1 - \lambda_t
(C_3+C_9)\right]  - M_{a\kt}^{P1} \lambda_t (C_5+C_7)- F_{a\kt}^{P2}
\lambda_t (a_6\right.\non && \left.+a_8) + F_{a\kt} \left[ \lambda_u
a_1 - \lambda_t (a_4+ a_{10}) \right] \right\} + \zeta_q \left\{
(f_{\kt}F_{e\eta_q} +F_{a\eta_q} ) \left[ \lambda_u
a_1\right.\right.\non && \left.\left. -\lambda_t (a_4+a_{10})\right]
-(\bar{f}_{\kt}F_{e\eta_q}^{P2}+F_{a\eta_q}^{P2}) \lambda_t (a_6 +
a_8) + (M_{e\eta_q} +M_{a\eta_q} ) \right.\non && \left.\cdot\left[
\lambda_u C_1 - \lambda_t (C_3+C_9)\right] -
(M_{e\eta_q}^{P1}+M_{a\eta_q}^{P1}) \lambda_t (C_5+C_7) \right\}\;,
\label{eq:ktpe} \eeq } where $\lambda_u= V^*_{ub} V_{us}$,
$\lambda_t= V^*_{tb} V_{ts}$ and $\zeta_{q(s)}=
\frac{\cos\phi}{\sqrt{2}}(-\sin\phi)$ for $\eta_q(\eta_s)$ with the
flavor mixing angle $\phi=39.3^\circ$. For $B^0 \to {\kt}^0 \eta$
decay, we find the similar result.

For $B \to {\kt} \eta'$ channels, the total decay amplitudes can be
easily obtained  by replacing $\zeta_{q(s)}$ with $\zeta'_{q(s)}=
\frac{\sin\phi}{\sqrt{2}}(\cos\phi)$ for $\eta'_q(\eta'_s)$ in
Eq.(\ref{eq:ktpe}).

\section{Numerical results and Discussions}\label{sec:n-d}

In this section, we will calculate the CP-averaged branching ratios
for those considered decay modes. The input parameters to be used
are given in Appendix \ref{sec:app2}. In numerical calculations, the
central values of input parameters will be used implicitly unless
otherwise stated.

Firstly, we find the pQCD predictions for the corresponding form
factors at zero momentum transfer: \beq F^{B\to \kt}_{0,1}(q^2=0)&=&
-0.44^{+0.06}_{-0.07}
(\omega_b)^{+0.04}_{-0.04}(\bar{f}_{\kt})^{+0.02}_{-0.02}(B_{1,3})\;,
\rm (Scenario\ \ \ I) \non F^{B\to \kt}_{0,1}(q^2=0)&=&
+0.76^{+0.12}_{-0.10}
(\omega_b)^{+0.08}_{-0.08}(\bar{f}_{\kt})^{+0.07}_{-0.07}(B_{1,3})\;,\rm
(Scenario\ \ \ II) \eeq for $f_B = 0.19$ GeV, and $\omega_b=0.40\pm
0.04$ GeV. They agree well with those as given in Ref.~\cite{LWL08}.

Using the decay amplitudes obtained in last section, it is
straightforward to calculate the branching ratios. The leading order
pQCD predictions for  the CP-averaged branching ratios in Scenario I
are the following (in unit of $10^{-6}$) \beq
 Br( B^+ \to {\kt}^+ \eta) &=& 11.8^{+5.3+0.3+1.1+2.5}_{-3.5-0.4-1.2-2.3} (19.2),
 \label{eq:bree1}\\
 Br( B^+ \to {\kt}^+\eta^{\prime}) &=& 21.6^{+1.6+3.1+4.0+4.5}_{-0.5-2.8-3.6-4.1}
(15.4), \label{eq:brep1} \\
 Br( B^0 \to {\kt}^0\eta) &=& 9.1^{+4.4+0.0+1.1+2.0}_{-2.8-0.1-1.1-1.8}
 (17.0), \label{eq:brpp1}\\
 Br(\ B^0 \to {\kt}^0\eta') &=& 22.0^{+1.6+3.2+3.9+4.6}_{-0.5-3.6-3.0-4.2}
(15.0),\label{eq:brkep1}
 \eeq
 and in Scenario II,
 \beq
 Br(\ B^+ \to {\kt}^+ \eta)&=&  33.8^{+13.5+1.1+7.7+8.2}_{-9.0-1.1-7.0-7.3}
 (38.8), \label{eq:bree2}\\
Br(\ B^+ \to {\kt}^+\eta^{\prime}) &=&
77.5^{+15.8+6.2+21.0+18.0}_{-10.8-5.8-16.5-16.1}
(49.6), \label{eq:brep2} \\
Br(\ B^0 \to {\kt}^0\eta)
&=&28.4^{+11.6+1.4+6.4+6.9}_{-7.8-1.4-5.9-6.2}
(34.2), \label{eq:brpp2}\\
 Br(\ B^0 \to {\kt}^0\eta') &=& 74.2^{+15.0+6.4+20.5+17.2}_{-10.3-5.7-16.2-15.5}
 (48.2),\label{eq:brkep2}
\eeq where the numbers in parentheses are the central values of
branching ratios without the inclusion of annihilation diagrams. The
first theoretical error is induced by the uncertainty of
$\omega_b=0.40 \pm 0.04$ GeV. The second uncertainty arises from the
Gegenbauer moment $a_2^{\etap} = 0.115 \pm 0.115$. The last two
errors are from the combinations of Gegenbauer coefficients $B_1$
and/or $B_3$ and decay constants $f_{\kt}$ and/or $\bar{f}_{\kt}$ of
the scalar meson $\kt$, respectively.

Now some phenomenological discussions are in order:
\begin{itemize}
\item[]{(1)}
In the evaluations of $B \to \kt \etap$ modes, the updated
parameters in the distribution amplitudes of $\etap$ mesons, for
example, $a_2^{\etap}=0.115 \pm 0.015$ and $a_4^{\etap}=-0.015$ were
used.

\item[]{(2)}
From the branching ratios for $B \to \kt \etap$ decays as shown in
Eq.~(\ref{eq:bree1}-\ref{eq:brkep2}), one can see that the results
in Scenario II are nearly 3-4 times large as those in Scenario I for
$B \to \kt \eta$ and $B \to \kt \eta'$ decays, respectively. This is
because the decay constants in Scenario II are larger and
sign-flipped, which results in the large branching ratios in
Scenario II. The pQCD predictions in Scenario I are preferable by
the existing data than in Scenario II.

\item[]{(3)}
As shown in Eq.(\ref{eq:bree1}-\ref{eq:brkep2}), the annihilation
diagrams play a more important role in contributing to the branching
ratios in Scenario I than that in Scenario II for $B \to \kt \eta$,
while the situation for $B \to \kt \eta'$ is quite the contrary. By
neglecting the annihilation contributions, for example, the pQCD
prediction for the central value of $Br(B^+ \to {\kt}^+ \eta)$ is
from $11.8 \times 10^{-6}$ to $19.2 \times 10^{-6}$ in Scenario I
while from $33.8 \times 10^{-6}$ to $38.8 \times 10^{-6}$ in
Scenario II; the prediction for $Br(B^+\to {\kt}^+ \eta^\prime)$
will change from $21.6\times 10^{-6}$ to $15.4\times 10^{-6}$ in
Scenario I, the corresponding change, however, is from $77.5\times
10^{-6}$ to $49.6\times 10^{-6}$ in Scenario II.

\item[]{(4)}
The long distance re-scattering effects may also affect the
branching ratios of $B \to S P$. We here do not consider such
effects since it is still very difficult to estimate them reliably
now.

\end{itemize}

It is worth mentioning that the authors of Ref.~\cite{CCY06,SWL07}
have studied four $B \to \kt \pi$ decays by employing the QCDF and
pQCD approach, respectively. They found that Scenario II is more
preferable than Scenario I by comparing with the data. But the
numerical results of branching ratios are very different in those
two papers.

We also performed the calculations for the four $B \to \kt \pi$
decays in the pQCD approach, and confirmed that the large branching
ratios could be obtained if the old Gegenbauer
moments~\cite{Shen09}, i.e., $a_2^\pi=0.44$ and $a_4^\pi=0.25$, are
used. By using the updated values of  $a_2^\pi=0.115 \pm 0.115$ and
$a_4^\pi=-0.015$, we find that the corresponding pQCD predictions
for the branching ratios of  $B \to \kt \pi$ decays are decreased
significantly by around 40\% (see Table~\ref{tab:brna}) in both
scenarios.

Frankly speaking,  the theoretical predictions for the branching ratios
of all considered $B \to SP$ decays still have a very large parameter-dependence.
we can not determine with enough confidence which scenario is the better one at present.
Much more theoretical studies and larger data sample are required to
understand the structure of scalar meson.

\begin{table}[htb]
\caption{The pQCD predictions for the branching ratios(in unit of $10^{-6}$) for $B \to \kt
\pi$ decays, obtained by using the new Gegenbauer moments in present work or the
old ones in Ref.~\cite{SWL07} respectively, where the various errors have been
added in quadrature. By comparison, we also cite the measured values
as given in~\cite{hfag08,pdg08}.}
\label{tab:brna}
\begin{center}\vspace{-0.5cm}
\begin{tabular}[t]{c|l|l|l|l|l} \hline  \hline
Modes  & Scenario I   & Scenario II & Scenario I~\cite{SWL07}   &
Scenario II~\cite{SWL07}  & Data
\\ \hline
  $B^+ \to {\kt}^+ \pi^0$   & $7.8^{+2.8}_{-2.3}$   & $21.6^{+8.5}_{-6.6}$& $11.3^{+2.5}_{-2.2}$   & $28.8^{+7.8}_{-7.3}$ & $-$\\
 $B^0 \to {\kt}^0 \pi^0$    & $5.8^{+1.7}_{-1.5}$   &$10.7^{+4.1}_{-3.2}$& $10.0^{+2.4}_{-2.2}$   &$18.4^{+6.1}_{-5.1}$ & $11.7^{+4.2}_{-3.8}$~\cite{hfag08} \\
 $B^+ \to {\kt}^0 \pi^+$    & $13.6^{+4.2}_{-3.6}$   &$30.9^{+12.4}_{-9.2}$& $20.7^{+4.7}_{-4.3}$   &$47.6^{+13.8}_{-11.9}$ & $47.0 \pm 5.0$~\cite{pdg08} \\
 $B^0 \to {\kt}^+ \pi^-$    & $13.2^{+4.0}_{-3.4}$   & $31.6^{+12.4}_{-9.3}$& $20.0^{+4.6}_{-4.2}$   & $43.0^{+12.8}_{-10.9}$ & $50.0^{+8.0}_{-9.0}$~\cite{pdg08} \\
 \hline \hline
\end{tabular}
\end{center}
\end{table}


In short, we calculated the branching ratios for $B \to \kt \etap$
decays by using the pQCD factorization approach at leading order. We
perform the evaluations in two scenarios for the scalar meson
spectrum. Besides the usual factorizable diagrams, the
non-factorizable and annihilation diagrams are also calculated
analytically in the pQCD approach. we find that (a) the pQCD
predictions for $Br(B \to \kst \eta^{(\prime)})$ which are about
$10^{-5}-10^{-6}$, basically agree with the data, but the
theoretical error is still large; (b) the agreement between the pQCD
predictions and the data in Scenario I is better than that in
Scenario II, which can be tested by the forthcoming LHC experiments;
(c) the annihilation contributions play an important role for these
considered decays. (d) much more theoretical studies and larger data
sample are needed to have a better understanding about the structure
of scalar mesons.

\begin{acknowledgments}

X.~Liu would like to thank Run-hui~Li for useful discussions.

\end{acknowledgments}


\begin{appendix}

\section{Input parameters and wave functions} \label{sec:app2}

The masses, decay constants, QCD scales  and $B$ meson lifetime used
in the calculations are
\beq
\Lambda_{\overline{\mathrm{MS}}}^{(f=4)} &=& 0.250 {\rm GeV}, \quad
f_\pi = 0.130 {\rm GeV}, \quad  f_{B} = 0.190 {\rm GeV}, \quad m_{\kt}=1.425 {\rm GeV}, \non
 m_0^{\eta_{q}}&=& 1.07 {\rm GeV},\quad m_0^{\eta_s}=1.92 {\rm
 GeV},  \quad
  \tau_{B^{\pm}}=1.638\times10^{-12}{\rm s},\quad M_W = 80.41{\rm
 GeV}, \non
 M_{B} &=& 5.28 {\rm GeV},  \quad \tau_{B^{0}}=1.53\times10^{-12}{\rm s}\nonumber.
 \label{para}
\eeq

For the CKM matrix elements, here we adopt the Wolfenstein
parametrization for the CKM matrix, and take $\lambda=0.2257,
A=0.814, \bar{\rho}=0.135$ and $\bar{\eta}=0.349$  \cite{pdg08}.

\end{appendix}


\end{document}